\documentclass[useAMS,usenatbib]{mn2e}

\usepackage{times}
\usepackage{graphicx}
\usepackage{amsmath}
\usepackage{amssymb}
\usepackage{natbib}
\usepackage{color}
\usepackage{epsfig}
\usepackage{psfig}
\usepackage{journal_names}
\usepackage{paralist}

\newcommand{\kms}{\mbox{\,km~s$^{-1}$}}

\newcommand{\comment}[1]{}

\title[Supermassive black hole connections with bulges and haloes]
{The SLUGGS survey: Probing the supermassive black hole connection with bulges and haloes using red and blue globular cluster systems}
\author[Pota~et~al.~ ]
{Vincenzo Pota$^{1}$, Alister W. Graham $^{1}$, Duncan A. Forbes$^{1}$,  Aaron J. Romanowsky$^{2, 3}$, \\
\\
\normalfont{\LARGE Jean P. Brodie$^{3}$,  Jay Strader$^{4}$}  \\ 
\\
$^1$ Centre for Astrophysics \& Supercomputing, Swinburne University, Hawthorn VIC 3122, Australia\\
$^2$ Department of Physics and Astronomy, San Jos\'e State University, One Washington Square, San Jose, CA 95192, USA\\
$^3$ University of California Observatories, 1156 High Street, Santa Cruz, CA 95064, USA\\
$^4$ Department of Physics and Astronomy, Michigan State University, East Lansing, Michigan 48824, USA\\
Email: vpota@astro.swin.edu.au, dforbes@astro.swin.edu.au}
 
\date{Released 2012 Xxxxx XX}

\pagerange{\pageref{firstpage}--\pageref{lastpage}} \pubyear{2012}

\begin{document}

\label{firstpage}

\maketitle

\begin{abstract}
Understanding whether the bulge or the halo provides the primary link to the growth of supermassive black holes has strong implications for galaxy evolution and supermassive black hole formation itself. 
In this paper, we approach this issue by investigating extragalactic globular cluster (GC) systems, which can be used to probe the physics of both the bulge and the halo of the host galaxy. 
We study the relation between the supermassive black hole masses $(M_{\rm BH})$ and the globular cluster system velocity dispersions $(\sigma_{\rm GC})$ using an updated and improved sample of 21 galaxies. 
We exploit the dichotomy of globular cluster system colours, to test if the blue and red globular clusters correlate differently with black hole mass. This may be expected if they trace the potentially
different formation history of the halo and of the bulge of the host galaxy respectively.  
We find that $M_{\rm BH}$ correlates with the total GC system velocity dispersion, although not as strongly as claimed by recent work of Sadoun \& Colin. 
We also examine the $M_{\rm BH} - \sigma_{\rm GC}$ relation for barred/bar-less and core/non-core galaxies, finding no significant difference, and for the first time we quantify
the impact of radial gradients in the GC system velocity dispersion profile on the $M_{\rm BH} - \sigma_{\rm GC}$ relation.
We additionally predict $M_{\rm BH}$  in 13 galaxies, including dwarf elliptical galaxies and the cD galaxy NGC~3311.
We conclude that our current results cannot discriminate between the bulge/halo scenario. Although there is a hint that the red GC velocity dispersion might correlate 
better with $M_{\rm BH}$ than the blue GC velocity dispersion, the number statistics are still too low to be certain. 

\end{abstract}


\section{Introduction}

Extragalactic globular clusters (GCs) may provide key insight into the connection between galaxies and supermassive black holes (SMBHs). 
GCs are typically old ($> 10$ Gyr, \citealt{Brodie}) and may have witnessed the events which formed the SMBH 
in the first place. Moreover, GC systems usually come in two subpopulations, thought to be the result of different formation mechanisms \citep[e.g.][]{Ashman92,Forbes97,Cote98}.
The blue (metal-poor) subpopulation has been associated with galaxy halos \citep{Forte,Moore,Forbes12,Spitler12}. It may have originated in metal-poor dwarf 
galaxies at high redshift consequently accreted into the halo of larger systems \citep{Elmegreen}. The properties of the red (metal-rich) GCs are 
similar to those of the galaxy bulge \citep{Strader11, Forbes2768}, perhaps because of a coeval formation, such as in a turbulent disk \citep{Shapiro} or in a merger \citep{Kruijssen12}. 
Therefore, if the growth of SMBHs is primarily driven by recent merger events, one might expect a stronger correlation between red GCs and SMBHs . 
Conversely, if the properties of SMBHs were set during the primordial formation of their host galaxies, we might expect a stronger correlation with blue GCs \citep{Omukai08,Mayer10,Debattista13}.

There exists a surprisingly good correlation between the total number of GCs  (both blue and red) per galaxy $(N_{\rm GC})$ and the black hole mass of galaxies $(M_{\rm BH})$.
However, this does not necessarily imply a primary correlation between GCs and SMBHs \citep{Jahnke11}. 
In fact, \citet{Snyder11} argued this correlation to be indirect as expected if it was a consequence of the debated black hole fundamental plane \citep{Hopkins07A,Graham08}. 
Nevertheless, the $M_{\rm BH} - N_{\rm GC}$ relation has been shown to have an intriguingly small scatter at fixed $M_{\rm BH}$ \citep{Burkert10,Harris11}. 
\citet{Rhode12} has recently shown that these findings are driven by low number statistics, and that an improved galaxy sample returns a scatter at fixed
$M_{\rm BH}$ which is larger than previously inferred. Rhode additionally found similar slopes and scatters for the relations for the blue and the red GCs.

Recently, \citet{Sadoun12} (hereafter SC12), have examined the correlation between the GC system velocity dispersion and $M_{\rm BH}$ for twelve 
galaxies, including the Milky Way. Their results suggest a tight correlation between $M_{\rm BH}$ and the velocity dispersion for both the red and blue GC subpopulations, 
with an intrinsic scatter $\epsilon$ always $\le 0.33$ dex, indicating a very tight correlation.
They also find that the red GCs are more closely correlated ($\epsilon=0.22$ dex) with $M_{\rm BH}$ than the blue GCs ($\epsilon=0.33~$dex). 

In this paper we revisit the work of SC12 with an expanded sample of 21 galaxies and updated $M_{\rm BH}$ values. We supplemented our sample with 
high velocity resolution data from the ongoing SLUGGS survey \citep{Pota13} and we re-analysed literature data with the same method. We tested if the tight 
correlation seen for the red GCs is real or driven by sample selection or methodology biases. The outline of the paper is as follows. We describe the data in 
Section \ref{sec:data} and their analysis in Section \ref{sec:method}. Results are then presented and discussed in Section \ref{sec:results}. Conclusions are given 
in Section \ref{sec:conclusions}.

\section{Galaxy sample}
\label{sec:data}

We study a subset of galaxies with direct $M_{\rm BH}$ measurements and with more than ten GC radial velocity measurements.
From the literature, we compiled a list of $13$ galaxies. This includes all the galaxies discussed in SC12, excluding the Milky Way, 
and two additional galaxies: NGC~$253$ \citep{Olsen} and NGC~$3585$ \citep{Puzia04}, not studied by SC12 because the uncertainties on the GC velocity 
dispersion were not quoted in the parent papers. 
The Milky Way is not included in this study because the results of \citet{Cote99} suggest that the still uncertain velocity dispersion of the Milky Way 
GC system is unusually large for its black hole mass. Moreover, the fact that the Milky Way GC analysis is carried out in three-dimensions
rather than in projection, makes the comparison with other galaxies not straightforward.  
We also update the GC catalogue used by SC12 for NGC~$4594$ with the latest compilation of \citet{Alves-Brito}.
We note that SC12 used $M_{\rm BH}$ values from \citet{Gultekin09b} although more recent $M_{\rm BH}$ were sometimes available.

For NGC~224 (M31) we use the GC system velocity dispersion measurements from \citet{Lee08Andromeda}, because their catalogue is not available on-line. 

In regard to NGC~253, there are two public GC catalogues for this galaxy: \citet{BeasleyN253} and \citet{Olsen}, for a total 38 GCs. 
However, we were unable to find a reliable calibration offset between the radial velocities of the four GCs in common between these two datasets. 
We decided to use the Olsen catalogue only, because it is larger in size (24 GCs) than Beasley's dataset (14~GCs).

The biggest strength of our data set is the addition of a further 9 new early-type galaxies from the SLUGGS survey, one of which (NGC~$4486$) was already discussed in SC12. 
We use the most recent black hole mass measurements as summarized in \citet{McConnell12} and \citet{Graham12}. 
This gives us a sample of 21 galaxies, nearly double the number used by SC12, which are listed in Table \ref{tab:data}.

\section{Method}
\label{sec:method}
\subsection{The globular cluster system velocity dispersion}
\label{sec:veldisp}
The stellar velocity dispersion, $\sigma_*$, used in the $M_{\rm BH} - \sigma_*$ relation is usually defined either as the luminosity-weighted velocity dispersion 
within $1/8^{\rm th}$ of an effective radius $R_e$, or within 1 $R_e$ $(\sigma_e)$, and/or as the central velocity dispersion $(\sigma_0)$. Although they represent 
physically distinct quantities, $\sigma_e$ and  $\sigma_0$ have been reported to be consistent with each other \citep{Gultekin09b}. 
This stems from the fact that the velocity dispersion profiles vary only weakly within these regions \citep[e.g.,][]{Emsellem11}.
\begin{figure}
\centering
\includegraphics[width=\columnwidth]{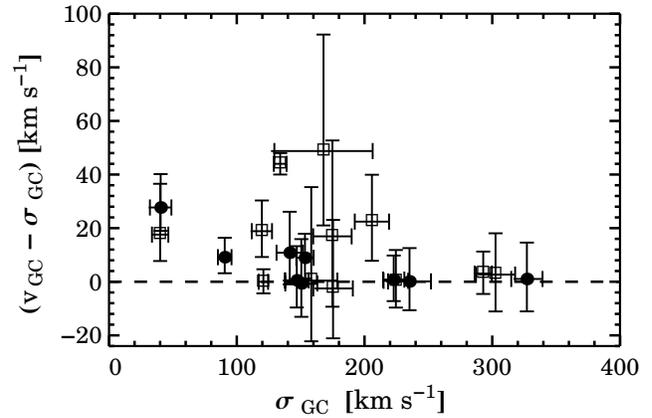} 
\caption{Difference between the rotation-subtracted velocity dispersion $\sigma_{\rm GC}$ and the rotation-included velocity dispersion $v_{\rm GC}$ 
without any colour split. Filled points and empty squares are the data from the SLUGGS survey and the literature respectively.
The two quantities are generally consistent with each other, but they disagree by up to $\sim 40$ km~s$^{-1}$ in systems with significant rotation.}
\label{fig:sigma}
\end{figure}

The detection of extragalactic globular clusters occurs predominantly at $R \ge R_e$. Therefore none of the stellar velocity dispersion quantities are directly 
recovered with GC data. We define the GC system velocity dispersion in two different ways, which are similar to the quantities used for stellar data. 
This also takes into account that some GC systems can have a rotation component which is as large as that of the random motions \citep{Beasley06}.

The first quantity, $\sigma_{\rm GC}$, assumes a Gaussian velocity distribution and it is defined as the standard deviation with respect to the model function \citep{Cote01}:
\begin{equation}
v(\theta) = v_{\rm sys}+v_{\rm rot} \sin(\theta_0 - \theta),
\label{eq:1}
\end{equation}
which measures the GC rotation amplitude $v_{\rm rot}$ as a function of the azimuth $\theta$, with $\theta_0$ being the direction of the angular momentum vector and $v_{\rm sys}$ being the systemic velocity 
of the host galaxy.
We use a variation on Equation \ref{eq:1}, originally designed by \citet{Krajnovic06} for IFU data-cubes and then extended to sparsely sampled data by \citet{Proctor}.
We then minimise a $\chi^2$ function (see \citealt{Bergond06}) to compute the best fit parameters $(v_{\rm rot}, \sigma_{\rm GC}, \theta_0)$. 
Uncertainties were derived by bootstrapping the sample 1000 times to derive 68 per cent confidence intervals. 
We will refer to the rotation-subtracted velocity dispersion of the \textit{red}, \textit{blue} and \textit{all} GCs as $\sigma_{\rm GC, R}$, $\sigma_{\rm GC, B}$, $\sigma_{\rm GC}$ respectively.

\begin{figure*}
\centering
\includegraphics[scale=0.64]{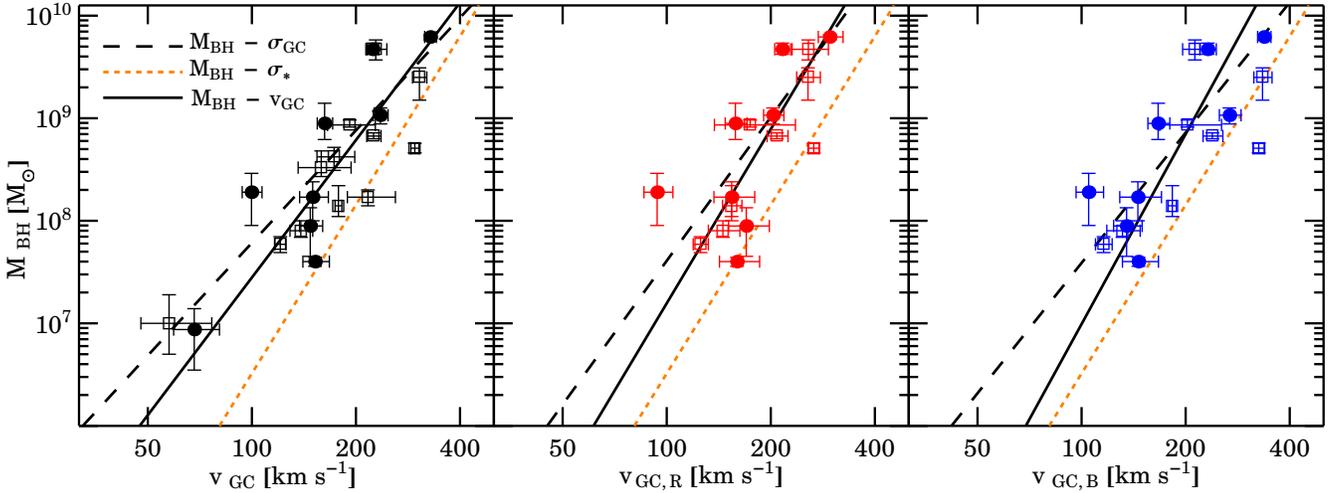} 
\caption{Black hole mass as a function of rotation-included GC system velocity dispersion. Left, central and right panels show the 
$M_{\rm BH} - v_{\rm GC}$ relation for all, red and blue GCs respectively. Data from the literature and from the SLUGGS 
survey are shown as open squares and filled points respectively. The black solid line is the best fit to the $M_{\rm BH} - v_{\rm GC}$ relation. 
The dashed lines are the best-fit to the $M_{\rm BH} - \sigma_{\rm GC}$ relations (whose datapoints are not plotted here for clarity). 
The slope and the intercept of the best-fit lines are the average between the values from the forward and inverse regression (see Table \ref{tab:results}).  
The dotted-orange line is the stellar $M_{\rm BH} - \sigma_*$ relation from the average between the forward and the inverse regression from \citet{Graham11}: 
$\alpha=8.14\pm0.05$ and $\beta=5.54\pm0.40$.}
\label{fig:vrms}
\end{figure*}

\begin{table*}
\centering
\label{mathmode}
\begin{tabular}{@{}l c c c c c c c c c c c c}
\hline
\multicolumn{1}{c}{}&\multicolumn{4}{c}{\textsc{Forward regression}}&\multicolumn{1}{c}{}&\multicolumn{4}{c}{\textsc{Inverse regression}}\\
\hline
Sample & N & $\alpha$ & $\beta$  & $\epsilon$ [dex]& $\Delta$& &$\alpha = - \alpha_{\rm inv} / \beta_{\rm inv}$ & $\beta = 1 /  \beta_{\rm inv}$& $\epsilon = \epsilon_{\rm inv} / \beta_{\rm inv}$ [dex] & $\Delta$  \\
\hline
\hline
$M_{\rm BH} - \sigma_{\rm GC}$& 21 & $8.76 _{-0.11} ^{+0.11}$ & $3.22 _{-0.33} ^{+0.48}$ & $0.42  _{-0.09} ^{+0.06}$ & 0.41 &  \vline& $8.86 _{-0.12} ^{+0.13}$ & $4.35 _{-0.61} ^{+1.02}$ & $0.48  _{-0.11} ^{+0.07}$ &0.51 \\ 
$(M_{\rm BH} >10^7 M_{\sun}) - v_{\rm GC}$ & 19 & $8.75 _{-0.11} ^{+0.11}$ & $3.85  _{-0.76} ^{+0.93}$ & $0.43  _{-0.06} ^{+0.10}$& 0.44& \vline& $8.80 _{-0.13} ^{+0.14}$ & $6.18 _{-0.97} ^{+1.37}$ & $0.54  _{-0.13} ^{+0.08}$ &0.57 \\
$M_{\rm BH} - v_{\rm GC}$ & 21 & $8.75 _{-0.10} ^{+0.11}$ & $3.74  _{-0.46} ^{+0.59}$ & $0.40  _{-0.06} ^{+0.09}$& 0.42& \vline& $8.83 _{-0.12} ^{+0.12}$ & $5.16 _{-0.62} ^{+0.91}$ & $0.47  _{-0.12} ^{+0.07}$ &0.52 \\
$(M_{\rm BH} >10^7 M_{\sun}) -  \sigma_{\rm GC}$ & 19 & $8.83 _{-0.10} ^{+0.12}$ & $3.59  _{-0.66} ^{+0.84}$ & $0.41  _{-0.06} ^{+0.10}$& 0.42& \vline& $8.93 _{-0.13} ^{+0.15}$ & $5.63 _{-0.88} ^{+1.18}$ & $0.52  _{-0.12} ^{+0.07}$ & 0.53 \\
$M_{\rm BH} -\sigma_*$ & 21 & $8.46 _{-0.10} ^{+0.07}$ & $4.44  _{-0.50} ^{+0.74}$ & $0.35  _{-0.05} ^{+0.08}$&0.37& \vline& $8.44 _{-0.11} ^{+0.08}$ & $5.48 _{-0.68} ^{+1.08}$ & $0.39  _{-0.09} ^{+0.06}$&0.42 \\
\hline
$M_{\rm BH} - \sigma_{\rm GC, B}$& 16 & $8.75 _{-0.12} ^{+0.14}$ & $3.45  _{-0.63} ^{+0.63}$ & $0.47  _{-0.12} ^{+0.07}$ &0.46& \vline& $8.82 _{-0.16} ^{+0.19}$ & $5.37 _{-0.86} ^{+1.14}$ & $0.58  _{-0.15} ^{+0.09}$ & 0.59 \\ 
$M_{\rm BH} - v_{\rm GC, B}$ &16 & $8.73 _{-0.11} ^{+0.13}$ & $3.50  _{-0.68} ^{+0.72}$ & $0.45  _{-0.12} ^{+0.07}$&0.45& \vline& $8.75 _{-0.15} ^{+0.16}$ & $5.53 _{-0.87} ^{+1.17}$ & $0.56  _{-0.16} ^{+0.09}$ &0.57 \\
\hline   
$M_{\rm BH} - \sigma_{\rm GC, R}$ &16 & $8.87 _{-0.12} ^{+0.14}$ & $3.77  _{-0.64} ^{+0.93}$ &  $0.47  _{-0.13} ^{+0.07}$& 0.47 & \vline& $9.02 _{-0.16} ^{+0.14}$ & $5.98 _{-1.07} ^{+1.51}$ & $0.60  _{-0.16} ^{+0.10}$ & 0.60 \\ 
$M_{\rm BH} - v_{\rm GC, R}$ &16 & $8.85 _{-0.12} ^{+0.12}$ & $4.50  _{-0.97} ^{+1.26}$ & $0.44  _{-0.13} ^{+0.07}$&0.47& \vline&$8.93 _{-0.14} ^{+0.15}$ & $6.77 _{-1.11} ^{+1.46}$ & $0.54  _{-0.16} ^{+0.09}$ &0.59 \\ 
\hline
\hline
\end{tabular}
\caption{Solutions to $\log (M_{\rm BH}/M_{\sun}) = \alpha+\beta \log (\sigma_{\rm GC}/200 \kms)$ for different GC subsets. Shown are the sample size $N$, the intercept $\alpha$, the slope $\beta$, the intrinsic
scatter $\epsilon$ and the total rms scatter $\Delta$ in the $\log M_{\rm BH}$ direction  for both the forward (minimise $\log M_{\rm BH}$ residual) and for the inverse regression (minimise $\log \sigma$ residual).}
\label{tab:results} 
\end{table*}

The second quantity, $v_{\rm GC}$, does not assume a Gaussian velocity distribution and it represents the azimuthally averaged second-order velocity moment which includes rotation:
\begin{equation}
v_{\rm GC}^2 =\frac{1}{N} \sum_{i=1}^N (v_{\rm i} - v_{\rm sys})^2 - (\Delta v_{\rm i})^2.
\label{eq:2}
\end{equation}
where $N$ is the sample size and $\Delta v_{\rm i}$ is the uncertainty on the radial velocity $v_i$ of the i$^{\rm th}$ globular cluster. The uncertainty 
on $v_{\rm GC}$ is estimated through the formula from \citet{Danese}. We will refer to $v_{\rm GC}$ of the \textit{red}, \textit{blue} and \textit{all} 
GCs as $v_{\rm GC, R}$, $v_{\rm GC, B}$, $v_{\rm GC}$ respectively.

The difference between $\sigma_{\rm GC}$ and  $v_{\rm GC}$ is that the former represents the rotation-subtracted velocity dispersion
whereas the latter also includes the rotation of the spheroid and it is a better reflection of specific kinetic energy.  A comparison between $\sigma_{\rm GC}$ and $v_{\rm GC}$ is given in 
Figure \ref{fig:sigma} for our galaxy sample without any GC subpopulation split. The two quantities are consistent with each other when the 
rotation component is negligible, as seen for several systems.

We perform a ``sanity check'' on all literature data. We prune GCs deviating more than $3\sigma$ from the local GC velocity distribution. We also clip 
outliers with unreasonably large uncertainty (usually $> 100$ km s$^{-1}$) and then we recalculate the respective $\sigma_{\rm GC}$ and $v_{\rm GC}$ 
to avoid methodology biases.

\subsection{The $M_{\rm BH} - \sigma_{\rm GC}$ and  $M_{\rm BH} - v_{\rm GC}$ relations for GC systems}
\label{sec:fitting}
Here we describe how we characterize the $M_{\rm BH} - \sigma_{\rm GC}$ relation. The procedure is identical for the $M_{\rm BH} - v_{\rm GC}$ relation.

In logarithmic space, $M_{\rm BH}$ and $\sigma_{\rm GC}$ appear to be linearly correlated. The relation we want to study is therefore:
\begin{equation}
\log  \left (\frac{M_{\rm BH}}{M_{\sun}} \right ) = \alpha + \beta  \log \left ( \frac{\sigma_{\rm GC}}{200\kms}\right),
\label{eq:3}
\end{equation}
where $\alpha$ and $\beta$ are the intercept and the slope of the relation. 
The numerical constant (200 km s$^{-1}$) is the normalization factor adopted in similar studies of the stellar $M_{\rm BH} -\sigma_*$ relation. 
We then use the $\chi^2$--minimization technique \citep{Press92} as modified by \citet{Tremaine02}. This ensures that the best fit to Equation~\ref{eq:3} is 
not biased in the case of large uncertainties \citep{Park12}.
Our minimization function is, using the notation~$y=\alpha+\beta x$:
\begin{equation}
\chi^2 (\alpha,\beta) \equiv \sum_{i=1}^N \frac{(y_i - \alpha - \beta x _i)^2}{\epsilon_{y, i} ^2 + \beta^2 \epsilon_{x, i}^2 + \epsilon^2}
\label{eq:4}
\end{equation}
where $\epsilon_x$ and $\epsilon_y$ are the errors on $x$ and $y$ respectively. These are defined as $\epsilon_x = (\log \sigma_{\rm upper} - \log \sigma_{\rm lower})/2$
and $\epsilon_y = (\log M_{\rm BH, \rm upper} - \log M_{\rm BH, \rm lower})/2$, respectively. 
The term $\epsilon$ is the intrinsic scatter in the $y$ direction in units of dex. $\epsilon$ is iteratively adjusted so that the value of $\chi^2 / (N-2)$ equals $1 \pm \sqrt{2/N}$.
Uncertainties on $\alpha$ and $\beta$ were obtained by bootstrapping the sample 2000 times and selecting the $68$ per cent confidence interval. 

This $\chi^2$ estimator does not treat the data symmetrically in the presence of intrinsic scatter. An ``inverse'' regression (minimizing the $\log \sigma_{\rm GC}$ residuals rather than
the $\log M_{\rm BH}$ residuals) can lead to very different slopes. The latter is preferable in the presence of possible Malmquist-type biases \citep[see][]{Graham11}. 
Given our ignorance of the physical mechanisms which links black hole mass to velocity dispersion,  there is no reason to believe that the forward regression should 
be favored over the inverse regression. Therefore, we perform both the ``forward'' and the ``inverse'' regression by replacing
$\epsilon$ in Equation \ref{eq:4} with $\beta^2 \epsilon^2$ as suggested by \citet{Novak06}. 

\section{results}
\label{sec:results}
The $M_{\rm BH}-v_{\rm GC}$ (and the $M_{\rm BH}-\sigma_{\rm GC}$) diagrams for our sample are shown in Figure~\ref{fig:vrms},
in which the final slope and intercept of the relations are the average between the forward and the inverse fit.
The respective best fit parameters are reported in Table \ref{tab:results}.

We find that $M_{\rm BH}$ correlates both with $\sigma_{\rm GC}$ and $v_{\rm GC}$ for all GC subsamples. However,  we note that the intrinsic scatter of all our GC 
subsets are at least two times larger than those reported by SC12. 
We find that this disagreement is driven by the $M_{\rm BH}$ values of five galaxies in the SC12 sample (marked in Table \ref{tab:data} with ``a'') for which we have updated $M_{\rm BH}$ measurements. 
In fact, reanalyzing the SC12 sample using our new velocity dispersion values and the $M_{\rm BH}$ values from SC12 (all from \citealt{GultekinA} and references therein), 
we always obtain $\epsilon \le 0.31$ dex, which is in agreement with their findings. Conversely, the regression on the SC12 sample using updated $M_{\rm BH}$ values, 
returns $\epsilon = 0.38$ dex for the full sample and $\epsilon = 0.44$ dex for the blue and red GC subsets. 
We conclude that the small intrinsic scatter of SC12 is driven by their black hole mass values and not by their GC system velocity dispersion data.  
This assumes that the latest values of $M_{\rm BH}$ that we adopt here are also more accurate than those which preceded them.
 
The slope, intercept and intrinsic scatter of the $M_{\rm BH} - v_{\rm GC}$ and the $M_{\rm BH} - \sigma_{\rm GC}$ relations are consistent with each other within the errors. 
Similarly, the differences found for the blue and red GCs are not statistically significant.
We note that the slopes of the $M_{\rm BH} - v_{\rm GC}$ relations are always steeper than the $M_{\rm BH}-\sigma_{\rm GC}$ ones, because $v_{\rm GC} > \sigma_{\rm GC}$ at low masses.
Also, the smaller intrinsic scatter with $v_{\rm GC}$ suggests that the GC kinetic energy (rotation plus dispersion) is a better predictor of black hole masses than 
the rotation-subtracted velocity dispersion. 

The intrinsic scatter of the $M_{\rm BH} - v_{\rm GC}$ and $M_{\rm BH} - \sigma_{\rm GC}$ relations are slightly larger than that of the stellar $M_{\rm BH}-\sigma_*$ relation 
from \citet{McConnell12} and \citet{Graham12}, who both find $\epsilon \sim 0.4$~dex. The best-fit to the stellar $M_{\rm BH}-\sigma_*$ relation computed using our 21 
galaxies has an intrinsic scatter of $\epsilon=0.35 _{-0.05} ^{+0.08}$~dex in the $\log M_{\rm BH}$ direction, which is also consistent with previous findings.

Lastly, it is noted that the stellar $M_{\rm BH} - \sigma_*$ relation in Figure \ref{fig:vrms} is shifted towards larger velocity dispersion values with respect to the $M_{\rm BH} -$ (GC system
velocity dispersion) relations. This offset is expected because $\sigma_*$ and the GC system velocity dispersion sample different regions, and maybe different physics, of the galaxy velocity dispersion profile. 
The stellar velocity dispersion, which probes $(R<R_e)$, is usually larger than the GC system velocity dispersion, which usually probes $R>R_e$. The difference $(\sigma_*-v_{\rm rms, A})$
is found to have a mean of ~ $35\pm6$ km s$^{-1}$ for our 21 galaxies.

\subsection{Radial trends}

It is interesting to see if the properties of the $M_{\rm BH} - v_{\rm GC}$ or $M_{\rm BH} - \sigma_{\rm GC}$ relation vary when the velocity dispersion is computed within different galactocentric radii.
\begin{figure}
\includegraphics[width=\columnwidth]{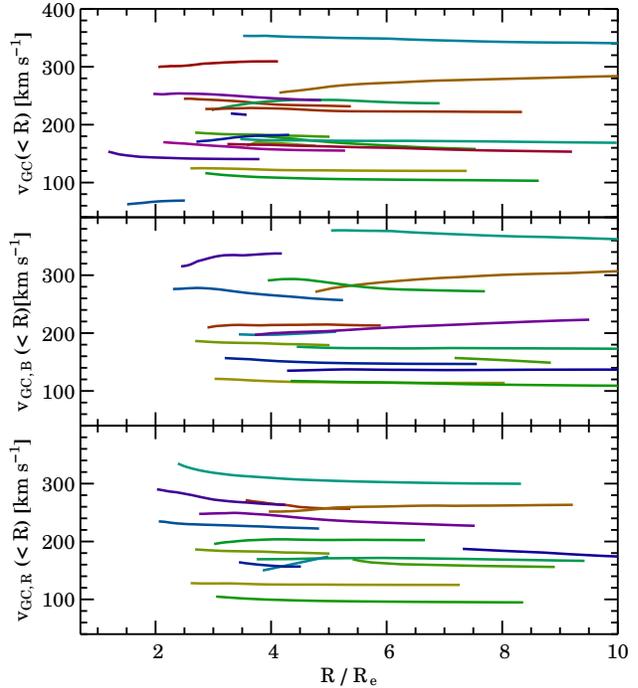} 
\caption{Cumulative root-mean-square velocity dispersion profiles. The plot shows the rotation-included velocity dispersion profiles within a certain radius for all (top panel), blue (central panel) and red GCs (bottom panel).
A running mean is used. Different colors represent different galaxies. Most of the profiles are generally flat at all radii.}
\label{fig:radialprofiles}
\end{figure}

To do so, we first normalize the galactocentric radii of each GC system to the host galaxy effective radius. We then perform $\chi^2$ tests (Equation \ref{eq:4}) with $v_{\rm GC}$ and 
$\sigma_{\rm GC}$ computed within different radial bins. For the sake of consistency, we adopt effective radii values from 2MASS, and we use the transformations from 
\citet{Cappellari11} to make them consistent with the values of the RC3 catalogue \citep{Vaucouleurs}. 

The cumulative velocity dispersion profiles for our galaxy sample are shown in Figure \ref{fig:radialprofiles} for all GC subsets. The profiles are generally flat over the radial range probed.
It is worth noting that GC dispersion profiles span different radial ranges depending on the galaxy, and we do not extrapolate the dispersion profiles to compensate for this effect.
Therefore, the number of GC systems within a given effective radius varies with the radius itself. 
Demanding a minimum of six GC systems per radial bin, we study the $M_{\rm BH} - v_{\rm GC}$ and the $M_{\rm BH} - \sigma_{\rm GC}$ relations between 3.5 and 5.5 $R_e$ 
for the blue and the red GC subpopulations.

\begin{figure}
\includegraphics[width=\columnwidth]{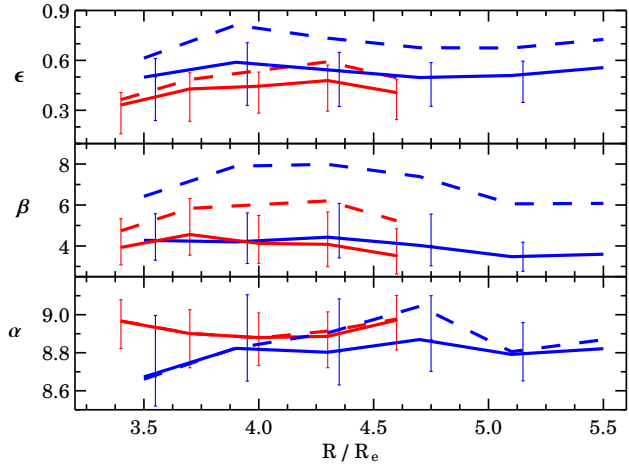} 
\caption{Best-fit $M_{\rm BH} - v_{\rm GC}$ relation within different radial bins. The plot shows how the best-fit $\alpha$, $\beta$ and $\epsilon$ vary when the $v_{\rm GC}$ is 
computed within an increasing number of effective radii. Blue and red colours represent the two GC subpopulations. Solid and dashed lines are the results from the forward and 
the inverse regressions respectively. The horizontal axis is the radius of the outermost GC in a given radial bin. For clarity, only the error bars from the forward regression are shown. 
None of the radial trends are statistically significant.}
\label{fig:radial}
\end{figure}

Results are shown in Figure \ref{fig:radial} for the $M_{\rm BH} - v_{\rm GC}$ relation. Each radial bin contains between six to a maximum of eleven GC systems. As expected from
the flatness of the velocity dispersion profiles (Figure \ref{fig:radialprofiles}), none of the radial trends seen in Figure \ref{fig:radial} are statistically significant. The relations for the blue and the red GC subpopulations 
are also statistically indistinguishable. 
There is an hint that the intrinsic scatter for the $M_{\rm BH} - v_{\rm GC, R}$ becomes smaller towards the central regions. This result is biased by the fact that the red GCs tend 
to be more centrally concentrated than the blue GGs. Given the small number statistics, the best fit to the $M_{\rm BH} - v_{\rm GC}$ relation is independent of radius within which 
the velocity dispersion is measured, at least for $R>R_e$. The same exercise performed on the rotation-subtracted velocity dispersion $\sigma_{\rm GC}$ leads to a similar result.

A caveat to bear in mind is the way the GC system velocity dispersion is computed. Ideally, one should weight the velocity dispersion for the 
GC surface density within a certain radius, similarly to what is done for the stellar velocity dispersion $\sigma_*$ (see Equation 1 in \citealt{McConnell12}). 
Similarly, the scale radius used in Figure \ref{fig:radialprofiles} should be the GC system's effective radius and not the host galaxy's effective radius.
However, GC surface density profiles are not available for all our galaxies. They are also dependent on variables such as GC selection criteria and imaging field-of-view,
which have been carried out differently in the literature. 

On the other hand, total GC system size scales with galaxy effective radius (Kartha et al. in prep.) and we see no strong variation of GC system velocity dispersion with radius.

\subsection{Cores and bars}

The stellar $M_{\rm BH}-\sigma_*$ relation is different for galaxies with or without bars \citep{Graham11}. It is thought that the orbital structure of the bar may elevate the apparent bulge 
velocity dispersion \citep{Bureau99}, resulting in an offset $M_{\rm BH}-\sigma_*$ relation for barred galaxies with the appropriate bar orientation. 
On the other hand, the $M_{\rm BH}-\sigma_*$ relation does not differ for non-barred galaxies with or without a `core' in the inner surface brightness profile \citep{Graham12}. 
An exception may however exist for ultramassive black holes such as those in NGC~3842 and NGC~4489 \citep{McConnell11a}. If these are included in the fit, the 
$M_{\rm BH}-\sigma_*$ relation for core galaxies is steeper $(\beta \sim 7.0)$ than that for non-core galaxies. 

We have tested if the trends seen for `core' and barred galaxies with stellar data are also present in our $M_{\rm BH} - v_{\rm GC}$ and $M_{\rm BH} - \sigma_{\rm GC}$ relations. 
To avoid low number statistics issues, we only look at the whole GC population, without any colour split. 
\begin{figure}
\centering
\includegraphics[width=\columnwidth]{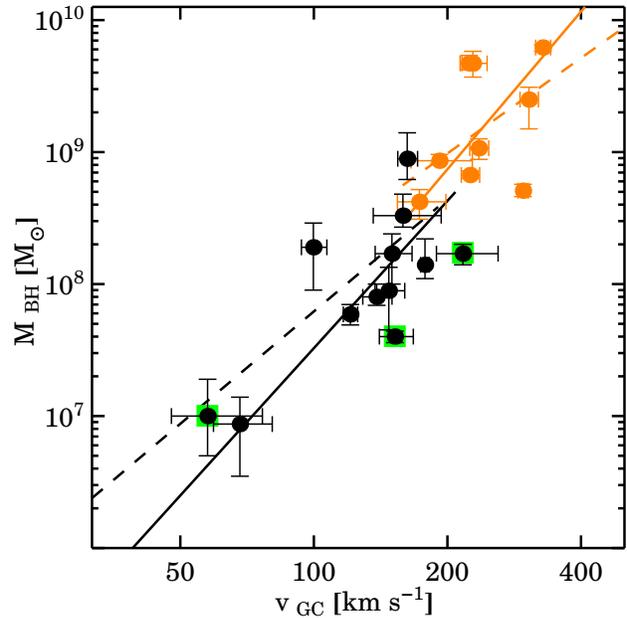} 
\caption{$M_{\rm BH} - v_{\rm GC}$ relation. Orange and black points are galaxies with and without a core in the 
inner surface brightness profile respectively. The filled and dashed lines are the best fits to core and non-core galaxies 
when using $v_{\rm GC}$ and $\sigma_{\rm GC}$ respectively. Green boxes mark barred galaxies (NGC~1316, NGC~1023 and NGC~253). 
The slope of the $M_{\rm BH} - v_{\rm GC}$ relation for core galaxies is consistent within the errors with that of non-core galaxies.}
\label{fig:core}
\end{figure}

Our sample contains only three barred galaxies (NGC~1023, NGC~1316 and NGC~253), preventing any statistical analysis. For the sake of completeness, we note that NGC~1023 
and NGC~1316 are indeed offset to higher velocity dispersions relative to the best-fit $M_{\rm BH} - v_{\rm GC}$ relation (Figure~\ref{fig:core}). However, only NGC~1023 is 
offset when considering $\sigma_{\rm GC}$. NGC~253 is neither offset from the $M_{\rm BH} - v_{\rm GC}$ nor the $M_{\rm BH} -\sigma_{\rm GC}$ relation, in agreement 
with what was found for stellar data. 

Regarding `core' galaxies, our sample contains nine core galaxies and twelve non-core galaxies (see Table \ref{tab:data}). 
The centre of the galaxy NGC~1407 is actually unclassified, but assume this galaxy to have a central core given its mass. 
We treat NGC~1316 (Fornax~A) as a cored galaxy \citep{Faber97}, but the reader should see the cautionary remarks 
in \citet{Graham12} regarding this galaxy's lack of a bulge/disc decomposition.

The relation between $M_{\rm BH}$ and GC system velocity dispersion for core/non-core galaxies is shown in Figure~\ref{fig:core}. We remind the reader that the 
final slope of the $M_{\rm BH}-$ (GC system velocity dispersion) relations is the average between the forward and the inverse regression. 
Using the uncertainty on the slope and intercept of each regression, we derived a weighted mean to account for the large uncertainties caused by low number statistics. For non-core galaxies, 
we obtain a slope of $\beta = 3.6 \pm 1.5$ and $\beta = 2.8 \pm 1.5$ for the $M_{\rm BH} -\sigma_{\rm GC}$ and $M_{\rm BH} - v_{\rm GC}$ relations respectively. 
For core galaxies, the uncertainty on the slope from the inverse regression is larger than the slope itself. This means that the final slope of this relation is driven only by 
that of the forward regression. In this case, we find $\beta = 2.2 \pm 1.6$ and $\beta = 2.4 \pm 1.6$ for the $M_{\rm BH} -\sigma_{\rm GC}$ and $M_{\rm BH} - v_{\rm GC}$ 
relations respectively. In conclusion, the $M_{\rm BH} - $ (GC system velocity dispersion) relations for core and non-core galaxies are consistent with each 
other as found by \citet{Graham12} with stellar velocity dispersion data.

\subsection{Predicting $M_{\rm BH}$ in other galaxies}

We exploit the best fit $M_{\rm BH} - $ (GC system velocity dispersion) relations found in this work to predict $M_{\rm BH}$ in galaxies without direct 
black hole mass measurements. We collected a sample of 13 galaxies with GC system kinematic information, listed in Table \ref{tab:predict}. The first 
four galaxies were re-analyzed in \citet{Pota13} with the methods described in \S \ref{sec:veldisp}. Similarly, we re-analyzed the GC system 
kinematics of NGC~4406 \citep{Park12GC} and of three luminous Virgo dwarf ellipticals (dEs) from \citet{Beasley09} and \citet{Beasley06}. 
Given that the $M_{\rm BH} - $ (GC system velocity dispersion) relations for the blue and the red GC subpopulations return consistent results, 
we decided to use the best fit $M_{\rm BH} - v_{\rm GC}$ relation:
\begin{equation}
\log  \left (\frac{M_{\rm BH}}{M_{\sun}} \right ) = 8.79+ 4.45  \log \left ( \frac{v_{\rm GC}}{200\kms}\right)
\label{eq:predict}
\end{equation}
where the slope and the intercept of this relation are the average between the forward and the inverse regression from Table \ref{tab:results}. 

Predicted black hole masses are given in Table \ref{tab:predict}. 
Particular emphasis should be given to the three Virgo dEs, whose predicted $M_{\rm BH}$ falls into the range of intermediate mass black holes ($\lesssim 10^6 M_{\sun}$). 
All three dEs are known to have a nuclear star cluster \citep{Ferrarese06}, whose masses are about one order of magnitude larger than our predicted black hole masses, as is expected \citep{Scott13}. 
In fact, the relation between the mass of the nuclear star cluster $M_{\rm NC}$ and stellar velocity dispersion $\sigma_*$ does not run parallel to the stellar $M_{\rm BH} - \sigma_*$. 
At fixed $\sigma_* \lesssim 150$ km s$^{-1}$, \citet{Graham12} shows that $M_{\rm NC} > M_{\rm BH}$, which is in agreement with our findings. 

It is also worth noting that NGC~3311, the dominant elliptical galaxy of the Hydra Cluster, is at first glance predicted to host an ultramassive black hole candidate with $M_{\rm BH} \sim 10^{11} M_{\sun}$. 
However, caveats here are the inclusion of ultra compact dwarfs (UCDs) which make up half of the kinematic sample of this galaxy. UCDs can be kinematically distinct from the underlying GC 
system \citep[e.g.][]{Strader11} and they can bias the velocity dispersion calculation. Another source of contamination might come from intra-cluster UCDs/GCs \citep{Misgeld,Richtler11}. 
Excluding the 52 UCDs and looking only at the GC sample, which may still be biased by the cluster potential, we obtain $M_{\rm BH} = 8.4 _{-2.6} ^{+4.7} \times 10^{10}$. 
This is still more massive than the most massive SMBH known today \citep{McConnell11a}.

\begin{table}
\centering
\label{mathmode}
\begin{tabular}{@{}l l l l l l}
\hline
Galaxy & Type &$v_{\rm GC}$ & $M_{\rm BH}$ & Ref. \\
 & & [km s$^{-1}$] &    [$M_{\sun}$]&   \\
\hline
\hline  
NGC~1380 &S0 &$160_{-17} ^{+23}$ & $2.2_{-0.9} ^{+1.8} \times 10^8$ & 1\\
NGC~3311 & cD&$653_{-40} ^{+48}$ & $1.2_{-0.3} ^{+0.4} \times 10^{11}$ & 2\\
NGC~3923 & E4&$273_{-29} ^{+42}$ & $2.4_{-0.9} ^{+2.1} \times 10^{9}$ & 3\\
NGC~4636 & E2&$212_{-10} ^{+11}$ & $7.9_{-1.5} ^{+2.0} \times 10^{8}$ & 4\\
NGC~4406 & E3&$295_{-36} ^{+54}$& $3.4_{-1.5} ^{+3.8} \times 10^{9}$ & 5\\
VCC~1261 & dE&$56_{-11} ^{+18}$ & $2.1_{-1.3} ^{+5.1} \times 10^{6}$& 6\\
VCC~1528 & dE&$52_{-15} ^{+22}$ & $1.5_{-1.1} ^{+5.7} \times 10^{6}$ & 7\\
VCC~1087 & dE&$41_{-10} ^{+14}$ & $5.2_{-3.7} ^{+14} \times 10^{5}$ & 8\\
\hline
NGC~1400 &S0 & $137_{-11} ^{+14}$& $1.1_{-0.3} ^{+0.6} \times 10^{8}$ & 9\\
NGC~2768 & E6&$165_{-11} ^{+13}$ & $2.6_{-0.6} ^{+1.0} \times 10^{8}$& 10\\
NGC~4278 &E2 &$177_{-7} ^{+9}$& $3.5_{-0.6} ^{+0.8} \times 10^{8}$ & 11\\
NGC~4365 & E3&$248_{-10} ^{+12}$& $1.6_{-0.2} ^{+0.3} \times 10^{9}$ & 12\\
NGC~4494 & E1&$99_{-12} ^{+14}$ & $2.6_{-1.1} ^{+2.1} \times 10^{7}$ & 13\\
\hline
\hline
\end{tabular}
\caption{Black hole mass predictions. Listed from the left to right are: galaxy name, morphological type, GC root-mean-square velocity dispersion, predicted black hole 
mass from Equation \ref{eq:predict} and the GC references.  Galaxies below the horizontal line are from the SLUGGS survey. 
References to GC data are: 1, \citet{Puzia04}; 2, \citet{Misgeld}; 3, \citet{Norris12}; 4, \citet{Lee10}; 5, \citet{Park12GC}; 6, 7, \citet{Beasley09}; 8, \citet{Beasley06}; 
9,10,11,12, \citet{Pota13};  13, \citet{Foster11}.}
\label{tab:predict} 
\end{table}

\section{Discussion and conclusions}
\label{sec:conclusions}
The aim of this paper was to test how well the velocity dispersion of extragalactic globular cluster systems correlates with the mass of supermassive black 
holes. This was motivated by the work of \citet{Sadoun12} who found an intriguingly tight correlation using 12 globular cluster systems. 

In this work we have extended this study to a sample of 21 GC systems and we have used the latest compilation of SMBH masses. 
We confirm that the velocity dispersion of GC systems correlates with SMBH mass. However, this correlation is less significant than 
that inferred by \citet{Sadoun12}. The tight correlation found by these authors was driven by old, and possibly less accurate, 
black hole mass values. We observe an rms scatter in excess of 0.4~dex in the $\log M_{\rm BH}$ direction.

We looked at the correlation between $M_{\rm BH}$ and the velocity dispersion of the blue and the red GC subpopulations separately. 
Different scatters are expected if blue and red GC systems trace the kinematics of the halo and the bulge of the host galaxy respectively. 
In the case of a stronger correlation with red GCs, this would suggest that SMBHs grew along with the stellar bulge. Conversely, a 
stronger correlation with blue GCs would suggest that SMBHs formation is more closely related with the halo. 
Our current results cannot discriminate between these two scenarios. In general, we find no significant difference between the $M_{\rm BH}-$ (GC system velocity 
dispersion) relation for the blue and the red GCs. This can be due to some factors discussed below. 

Ideally, one should analyze the bluer and the redder GCs for each GC system to avoid contamination in proximity to the blue/red dividing colour. This can make a 
difference in the final value of the GC system velocity dispersion \citep{Pota13}. At the same time, this would decrease the number statistics for most of the galaxies. 
Also, uneven GC spatial sampling can affect the final kinematic outcome, as seen for NGC~4636 in \citet{Schuberth12}. 

We have looked at the $M_{\rm BH}-$ (GC system velocity dispersion) relation computing the GC system velocity dispersion within different galactocentric radii, obtaining
no significant trends with radius. Collectively, this suggests either that the $M_{\rm BH}-$ (GC system velocity dispersion) relation is secondary, 
or that a larger galaxy sample will be needed to discriminate which of the GC subpopulation trends is the stronger.

We have looked for possible trends in the  $M_{\rm BH}-$ (GC system velocity dispersion) relation for core/non-core galaxies, 
finding similar slopes, in agreement with stellar velocity dispersion results \citep{Graham12}.

The best fit relation between $M_{\rm BH}$ and the rotation-included GC system velocity dispersion has been used to predict black hole masses in 13 galaxies. 
This implies that NGC~3311 contains an ultramassive black hole with $M_{\rm BH} \sim 10^{11} M_{\sun}$. 

\section*{Acknowledgements}

We thank the anonymous referee for the constructive feedback.
Some of the data presented herein were obtained at the W. M. Keck Observatory, operated as a scientific partnership among the California Institute of Technology, the 
University of California and the National Aeronautics and Space Administration, and made possible by the generous financial support of the W. M. Keck Foundation. The authors wish to recognize and acknowledge the 
very significant cultural role and reverence that the summit of Mauna Kea has always had within the indigenous Hawaiian community. The analysis pipeline used to reduce the DEIMOS data was developed at UC 
Berkeley with support from NSF grant AST-0071048. This work is based in part on data collected at Subaru Telescope and obtained from the SMOKA (which is operated by the Astronomy Data Centre, National 
Astronomical Observatory of Japan), via a Gemini Observatory time exchange. 
AWG is supported by the Australia Research Council (DP110103509 and FT110100263). This work was supported by NSF grants AST-0909037 and AST-1211995.
This research has made use of the NASA/IPAC Extragalactic Database (NED) which is operated by the Jet Propulsion Laboratory, California Institute of Technology, 
under contract with the National Aeronautics and Space Administration. We acknowledge the usage of the HyperLeda database (http://leda.univ-lyon1.fr).
\bibliographystyle{mn2e}
\bibliography{paper}
\appendix
\section{data table}
\label{appendix}

\begin{table*}
\centering
\label{mathmode}
\begin{tabular}{@{}l l l l l l l l l l l l l}
\hline
Galaxy & Type & $D$ & $M_{\rm BH}$ & $\sigma_*$  & Core & $\sigma_A$ & $\sigma_B$ &  $\sigma_R$ & $v_{\rm GC}$ & $v_{\rm GC, B}$ &  $v_{\rm GC, R}$ & Ref.   \\
  $[$NGC$]$  &  &     [Mpc] &   [$10^8 M_{\sun}$] &  [km s$^{-1}$] & &  [km s$^{-1}$]       &    [km s$^{-1}$]        &      [km s$^{-1}$ &      [km s$^{-1}$] &      [km s$^{-1}$] &      [km s$^{-1}$] &    \\
 (1)  & (2)      &  (3)            & (4)            & (5)     & (6)    &  (7)         &  (8)      &       (9)           &   (10) & (11)& (12) &(13)    \\
\hline
\hline  
0224& Sb & 0.73 & $1.4_{-0.3} ^{+0.8}$ 	& $160 \pm 8$	& n & $134_{-5} ^{+5}$ 	& $129_{-6} ^{+8}$ &	 $121_{-10} ^{+9}$ 	& $178_{-4} ^{+4}$ 	& $183_{-5} ^{+5}$ &	 $154_{-9} ^{+11}$ &	1 \\
0253	& SBc & 3.5&  $0.1_{-0.1} ^{+0.05}$ 	& $109 \pm 10$& n & $37_{-7} ^{+6}$ 	& $-$ &	 $-$ 	& $58_{-13} ^{+16}$ 	& $-$ &	 $-$ &	2 \\
0524	& S0& 24.2&  $8.6_{-0.4} ^{+1.0}$ 	& $235 \pm 10$& y	& $175_{-15} ^{+15}$ 	& $167_{-27} ^{+23}$ &	 $164_{-33} ^{+27}$ 	& $192_{-26} ^{+36}$ 	& $202_{-34} ^{+52}$ &	 $174_{-37} ^{+61}$ &	3 \\
1316	& SB0& 21.0 & $1.7_{-0.3} ^{+0.3}$ 	& $226 \pm 11$& y	& $168_{-41} ^{+38}$ 	& $-$ &	 $-$ 	& $217_{-28} ^{+43}$ 	& $-$ &	 $-$ &	4 \\
1399	& E1& 19.4&  $4.7_{-0.6} ^{+0.6}$ 	& $296 \pm 15$&y	& $293_{-8} ^{+7}$ 	& $321_{-12} ^{+12}$ &	 $269_{-10} ^{+9}$ 	& $296_{-8} ^{+8}$ 	& $325_{-12} ^{+13}$ &	 $266_{-9} ^{+10}$ &	5 \\
3031	& Sab & 4.1&  $0.9_{-0.11} ^{+0.20}$ 	& $143 \pm 7$     & n	& $120_{-9} ^{+8}$ 	& $124_{-13} ^{+13}$ &	 $99_{-8} ^{+8}$ 	& $139_{-10} ^{+11}$ 	& $131_{-13} ^{+16}$ &	 $145_{-13} ^{+17}$ &	6 \\
3379$^{\rm a}$	& E1& 10.7&  $4.2_{-1.1} ^{+1.0}$ 	& $206 \pm 10$& y	& $184_{-14} ^{+15}$ 	& $-$ &	 $-$ 	& $173_{-19} ^{+26}$ 	& $-$ &	 $-$ &	7/8 \\
3585	& S0& 20.6&  $3.3_{-0.6} ^{+1.5}$ 	& $213 \pm 10$& n	& $158_{-22} ^{+20}$ 	& $-$ &	 $-$ 	& $159_{-23} ^{+35}$ 	& $-$ &	 $-$&	9 \\
4472$^{\rm a}$	&E2 & 16.7&  $25_{-1} ^{+6}$ 	& $315 \pm 16$&y	& $303_{-13} ^{+15}$ 	& $333_{-20} ^{+19}$ &	 $261_{-18} ^{+18}$ 	& $305_{-14} ^{+15}$ 	& $334_{-19} ^{+22}$ &	 $256_{-19} ^{+22}$ &	10 \\
4594$^{\rm a}$	 &Sa& 10.0&  $6.7_{-0.4} ^{+0.5}$ 	& $230 \pm 12$&y	& $229_{-10} ^{+10}$ 	& $238_{-14} ^{+13}$ &	 $208_{-13} ^{+13}$ 	& $225_{-10} ^{+11}$ 	& $239_{-14} ^{+17}$ &	 $208_{-13} ^{+16}$ &	11 \\
4649$^{\rm a}$	&E2& 16.5&  $47_{-10} ^{+11}$ 	&$335 \pm 17$	& y & $206_{-13} ^{+13}$ 	& $194_{-16} ^{+14}$ &	 $228_{-28} ^{+28}$ 	& $228_{-15} ^{+17}$ 	& $213_{-17} ^{+21}$ &	 $257_{-26} ^{+37}$ &	12 \\
5128$^{\rm a}$	&S0& 4.1&  $0.59_{-0.10} ^{+0.11}$ 	& $150 \pm 7$	& n & $121_{-4} ^{+4}$ 	& $118_{-5} ^{+5}$ &	 $123_{-6} ^{+6}$ 	& $121_{-4} ^{+5}$ 	& $116_{-6} ^{+7}$ &	 $125_{-6} ^{+7}$ &	13 \\
\hline
0821$^{\rm a}$	&E6& 23.4&  $1.7_{-0.7} ^{+0.7}$ 	& $209 \pm 10 $ &n	& $151_{-13} ^{+13}$ 	& $129_{-20} ^{+19}$ &	 $162_{-19} ^{+20}$ 	& $150_{-12} ^{+17}$ 	& $145_{-17} ^{+25}$ &	 $154_{-17} ^{+26}$ & 14 \\
1023 & SB0&10.5&  $0.4_{-0.04} ^{+0.04}$ 	& $204 \pm 10$ &n	& $141_{-10} ^{+10}$ 	& $139_{-16} ^{+15}$ &	 $139_{-18} ^{+16}$ 	& $152_{-12} ^{+15}$ 	& $146_{-15} ^{+21}$ &	 $160_{-18} ^{+26}$ & 18\\
1407 & E0& 29.0 & $47_{-5} ^{+7}$ & $274 \pm 14$ & y & $222_{-8} ^{+8}$ & $231_{-11} ^{+11}$ &	 $210_{-10} ^{+11}$ & $223_{-8} ^{+9}$ 	& $231_{-11} ^{+13}$ &	 $215_{-10} ^{+12}$ &	 14\\
3115	&  S0 &9.5&  $8.9_{-2.7} ^{+5.1}$ 	& $230 \pm 11$ &n	& $153_{-7} ^{+7}$ 	& $152_{-8} ^{+9}$ &	 $150_{-10} ^{+10}$ 	& $162_{-8} ^{+9}$ 	& $166_{-11} ^{+13}$ &	 $158_{-10} ^{+13}$ & 15\\
3377$^{\rm a}$	& E5& 11.0&  $1.9_{-1.0} ^{+1.0}$ 	& $145 \pm 7$ &n	& $91_{-6} ^{+5}$ 	& $99_{-7} ^{+7}$ &	 $78_{-8} ^{+8}$ 	& $100_{-6} ^{+7}$ 	& $105_{-8} ^{+11}$ &	 $94_{-8} ^{+10}$ & 14 \\
4473$^{\rm a}$	& E5& 15.2&  $0.89_{-0.44} ^{+0.45}$ 	& $190 \pm 9$ & n & $147_{-9} ^{+9}$ 	& $134_{-11} ^{+11}$ &	 $162_{-16} ^{+14}$ 	& $148_{-10} ^{+13}$ 	& $135_{-11} ^{+15}$ &	 $170_{-19} ^{+28}$ & 16\\
4486$^{\rm a}$	& E1& 16.7&  $62_{-4} ^{+3}$ 	& $334\pm10$ & y & $327_{-10} ^{+12}$ 	& $336_{-17} ^{+16}$ &	 $293_{-25} ^{+27}$ 	& $328_{-12} ^{+14}$ 	& $337_{-14} ^{+16}$ &	 $296_{-22} ^{+28}$ & 17 \\
5846	& E0& 24.2&  $10.7_{-1.9} ^{+1.9}$ 	& $237 \pm 10$ & y & $235_{-18} ^{+17}$ 	& $269_{-18} ^{+17}$ &	 $201_{-13} ^{+13}$ 	& $235_{-11} ^{+12}$ 	& $268_{-17} ^{+21}$ &	 $203_{-12} ^{+15}$& 14 \\
7457$^{\rm a}$	& S0 &12.2&  $0.087_{-0.052} ^{+0.052}$ 	& $67 \pm 3$ & n	& $40_{-9} ^{+8}$ 	& $-$ &	 $-$ 	& $68_{-9} ^{+12}$ 	& $-$ &	 $-$ & 14 \\
\hline
\hline
\end{tabular}
\caption{Galaxy sample. 
Galaxy NGC names (1) and Hubble types (2) are from NED database. 
Galaxy distances (3), SMBH masses (4) and stellar velocity dispersions (5) are from \citet{McConnell12} and references therein. If not in \citet{McConnell12}: distances were obtained by 
subtracting 0.06 mag \citep{Mei} to the distance modulus from \citet{Tonry01}; central stellar velocity dispersions are weighted values from HyperLeda; $M_{\rm BH}$ are from \citet{Oliva95} and
\citet{Hu08} for NGC~253 and NGC~5846 respectively. (6) is the presence of a core in the galaxy inner surface brightness profile.
Column (7), (8) and (9) are the rotation-subtracted velocity dispersion for all, blue and red GCs respectively. 
Column (10), (11) and (12) are the root-mean square velocity for all, blue and red GCs respectively. 
GC references (13) : 1, \citet{Lee08Andromeda}; 2, \citet{Olsen}; 3, \citet{Beasley04N524}; 4, \citet{Goudfrooij01}; 5, \citet{Schuberth}; 6, \citet{Nantais10}; 7, \citep{Pierce06};
8, \citep{Bergond06}; 9, \citet{Puzia04}; 10, \citet{Cote03}; 11, \citet{Alves-Brito}; 12, \citet{Hwang}; 13, \citet{Woodley10}; 14, \citet{Pota13}; 15, \citet{Arnold}; 16, Foster et al. (in prep.);
17, \citet{Strader11}; 18, Pota et al. in prep. GC system velocity dispersion values for NGC~$224$ are from \citet{Lee08Andromeda}. Galaxies with updated $M_{\rm BH}$ measurements after \citet{GultekinA} are marked with (a).}
\label{tab:data} 
\end{table*}

\end{document}